\documentclass[12pt]{article}
\usepackage[T1]{fontenc}
\usepackage{graphicx}
\usepackage{amsmath}
\usepackage{amsfonts}
\usepackage[final]{changes}
\usepackage{enumerate}
\usepackage{authblk}
\usepackage{mathtools, nccmath}
\usepackage{url} 
\usepackage{natbib}
\usepackage{xcolor}
\usepackage{adjustbox}
\usepackage{rotating}

\newtheorem{theorem}{Theorem}
\newtheorem{definition}{Definition}
\newtheorem{example}{Example}
\title{Obtaining $(\epsilon,\delta)$-differential privacy guarantees when using a Poisson mechanism to synthesize contingency tables}
\author[$\dagger$]{James Jackson}
\author[$\star$]{Robin Mitra}
\author[$\dagger$]{Brian Francis}
\author[$\ddag$]{Iain Dove}
\affil[$\dagger$]{Lancaster University, Lancaster, UK}
\affil[$\star$]{University College London, London, UK}
\date{}
\affil[$\ddag$]{Office for National Statistics, Titchfield, UK}
\begin{document}
\maketitle              
\begin{abstract}
We show that differential privacy type guarantees can be obtained when using a Poisson synthesis mechanism to protect counts in contingency tables. Specifically, we show how to obtain $(\epsilon, \delta)$-probabilistic differential privacy guarantees via the Poisson distribution's cumulative distribution function. We demonstrate this empirically with the synthesis of an administrative-type confidential database.
\end{abstract}

\section{Introduction}
Differential privacy (DP) \citep{Dwork2006} is a property of a perturbation mechanism that formally quantifies how accurately any individual's true values can be established, given all other individuals' true values are known. Originally developed as a way to protect the privacy of summary statistics (queries), it soon expanded as a way to protect entire data sets. Differentially private data synthesis (DIPS) has since become a popular area of research; see, for example, \cite{Abowd2008, machanavajjhala2008privacy, Charest2011, McClure2012, Bowen2020, Quick2021, Drechsler2023}.

In \cite{Jackson2021, Jackson2022}, we proposed a synthesis approach for \replaced{contingency tables}{categorical data sets, which takes place at the tabular level, and} that uses saturated count models. This approach effectively uses a count distribution to apply noise to the counts in the original data's contingency table, and therefore shares traits with DP mechanisms which apply noise in a similar way. \added{Note that as microdata composed entirely of categorical variables can be expressed in contingency table format, this approach is suitable in the case of categorical data more generally.}
 
 In this paper, we consider the ability to obtain DP-guarantees when using the Poisson distribution to synthesize counts in \replaced{contingency tables}{tabular data (contingency tables)}. We show that although $\epsilon$-DP cannot be satisfied, $(\epsilon, \delta)$-DP guarantees can be obtained through the use of the Poisson's cumulative distribution function (CDF). 
 
 \added{The motivation behind this work is that, with the exception of \cite{Quick2021}, the use of count distributions has largely been overlooked as a way to satisfy DP. An obvious benefit of using count distributions is that negative counts cannot be obtained. As the Poisson has only one parameter and hence is likely to be sub-optimal, the intention is that in the future the Poisson could be replaced with more complex count distributions, such as the (discretised) gamma family distribution, where additional parameters provide scope for fine-tuning.} \par The paper is structured as follows. Section \ref{sec2} introduces some terminology and definitions. Section \ref{sec3} looks at existing DP mechanisms for contingency tables, such as the (discretised) Laplace and Gaussian mechanisms. Section \ref{sec4} gives our novel contribution, the ability to obtain $(\epsilon, \delta)$-DP guarantees when using a Poisson synthesis mechanism. Section \ref{sec5} gives an empirical example using an administrative database. Section \ref{sec6} gives some concluding remarks.
\section{Terminology and definitions} \label{sec2}
\cite{Rinott2018} set out how DP extends into a contingency table setting. Following their notation, let $\mathbf{a}=(a_k, \hdots, a_K) \in \mathcal{A}$ and $\mathbf{b}=(b_k, \hdots, b_K) \in \mathcal{B}$ denote vectors of \deleted{cell }counts in the original and synthetic data's contingency tables, respectively, where $K$ denotes the number of cells and $\mathcal{A}$ and $\mathcal{B}$ denote the range of obtainable original and synthetic counts (respectively). For contingency tables, we suppose that $\mathcal{A}=\mathcal{B}=\mathbb{Z}_{\geq 0}^K$, where $\mathbb{Z}_{\geq 0}$ is the set of non-negative integers. \par Moreover, we describe $\mathbf{a}$ and $\mathbf{a}^\prime$ as {neighbours}, denoted by $\mathbf{a} \sim \mathbf{a}^\prime$, whenever all but one of the counts in $\mathbf{a}$ and $\mathbf{a}^\prime$ are identical and the differing count differs by exactly one. Henceforth, without loss of generality, we suppose $\mathbf{a}$ and $\mathbf{a}^\prime$ differ in their $k$th element only, i.e.\ $a_k^\prime = {a}_k -1$ and $a_i=a_i^\prime$ for $i= 1, \hdots,K$, $i \neq k$. Thus $\mathbf{a}^\prime$ represents the data held by the intruder (who knows all but one of the individuals' true values) and $\mathbf{a}$ represents the completed data where the ``unknown individual'' has been added to the cell in which they truly belong. \par 
The $\epsilon$-DP definition revolves around the likelihood ratio, or, more accurately, around a series of likelihood ratios. 
\begin{definition}[$\epsilon$-DP] \label{def1}
A perturbation mechanism $\mathcal{M}$ satisfies $\epsilon$-DP ($\epsilon>0$) if: \begin{align}
\exp{(-\epsilon)} \leq \frac{\mathbb{P}\left(\mathcal{M}(\mathbf{a})=\mathbf{b}\right)}{\mathbb{P}\left(\mathcal{M}(\mathbf{a^\prime})=\mathbf{b}\right)}  \leq \exp{(\epsilon)}, \label{DP} \\  \quad \forall \; \mathbf{a} \sim \mathbf{a}^\prime \in \mathcal{A}, \; \forall \; \mathbf{b} \in \mathcal{B}. \nonumber 
\end{align}
\end{definition}
Definition \ref{def1} is the special case of the standard DP definition, given in \cite{Dwork2006}, for when the range of $\mathcal{A}$ and $\mathcal{B}$ are discrete. \added{Although we appreciate that in some instances the denominator in (\ref{DP}) could be equal to zero, for the mechanisms we consider here this probability is always non-zero.} 

For any $\mathbf{a}$, $\mathbf{a}^\prime$ and $\mathbf{b}$, whenever the ratio ${\mathbb{P}(\mathcal{M}(\mathbf{a})=b)}/{\mathbb{P}(\mathcal{M}(\mathbf{a^\prime})=b)}$ is either small or large, relatively too much is gleaned about the unknown individual's true values. It is worth noting, too, that the above definition considers all possible synthetic data sets in $\mathcal{B}$, illustrating that DP is not a risk metric for a particular synthetic data set but rather a property of a synthesis mechanism.  \par
Somewhat confusingly, there are two similar but different relaxations of $\epsilon$-DP. The first is $(\epsilon, \delta)$-differential privacy \citep{Dwork2014}. The second is known as $(\epsilon, \delta)$-probabilistic differential privacy \citep{machanavajjhala2008privacy}. These are given below in Definitions \ref{def2} and \ref{def3}. In the remainder of this paper, we focus on $(\epsilon, \delta)$-probabilistic DP. Yet whenever $(\epsilon, \delta)$-probabilistic DP is satisfied, $(\epsilon, \delta)$-DP is also satisfied \citep{Goetz2012}.  
\begin{definition}[$(\epsilon, \delta)$-DP] \label{def2}
A perturbation mechanism $\mathcal{M}$ satisfies $(\epsilon, \delta)$-DP ($\epsilon>0$; $0 \leq \delta \leq1$) if: \begin{align}
\frac{\mathbb{P}(\mathcal{M}(\mathbf{a})=b)- \delta }{\mathbb{P}(\mathcal{M}(\mathbf{a^\prime})=b)}&\leq \exp{(\epsilon)} \quad \text{and} \quad \frac{\mathbb{P}(\mathcal{M}(\mathbf{a}^\prime)=b)- \delta }{\mathbb{P}(\mathcal{M}(\mathbf{a})=b)} \leq \exp{(\epsilon)} \\  \quad \forall \; \mathbf{a} \sim \mathbf{a}^\prime \in \mathcal{A},\; \mathbf{b} \in \mathcal{B}. \nonumber
\end{align}
\end{definition}
 \begin{definition}[$(\epsilon, \delta)$-probabilistic DP] \label{def3}
A perturbation mechanism $\mathcal{M}$ satisfies $(\epsilon, \delta)$-\textit{probabilistic} DP ($\epsilon>0$; $0 \leq \delta \leq1$) if: 
\begin{align}
\mathbb{P} \bigg[ \frac{1}{\exp{(\epsilon)}} &\leq \frac{\mathbb{P}(\mathcal{M}(\mathbf{a})=b)}{\mathbb{P}(\mathcal{M}(\mathbf{a^\prime})=b)}\leq \exp{(\epsilon)} \bigg] > 1-\delta  \quad \forall \; \mathbf{a} \sim \mathbf{a}^\prime \in \mathcal{A},\; \mathbf{b} \in \mathcal{B}.
\end{align}
\end{definition}
\begin{theorem}[$(\epsilon, \delta)$-probabilistic DP implies $(\epsilon, \delta)$-DP] \label{the1}
If a perturbation mechanism $\mathcal{M}$ satisfies $(\epsilon, \delta)$-probabilistic DP, then it also satisfies $(\epsilon, \delta)$-DP. (Proof: see \cite{Goetz2012})
\end{theorem}
\section{Examples of existing DP mechanisms} \label{sec3}
We now give examples of existing DP mechanisms suitable for synthesizing counts in contingency tables. \added{Note that for the Laplace and Gaussian mechanisms, discretised noise needs to be added (unless one is willing to accept non-integer ``counts''). This can simply involve adding continuous noise before rounding the adjusted values to the nearest integer. Similarly, negative values can be rounded to zero. }
\begin{example}[The Laplace mechanism] \end{example}
A random variable $X \sim$ Laplace$(\mu ,d)$ has probability density function $f_L$:

\begin{align*}
f_L(x ; \mu ,d)&={\frac {1}{2d}}\exp \left(-{\frac {|x-\mu |}{d}}\right).  
\end{align*}
 The Laplace mechanism $\mathcal{M}_L$ satisfies $\epsilon$-DP by using the Laplace distribution to add random noise to the original counts $\mathbf{a}$. \added{Specifically, for every original count $a_i$, the Laplace mechanism generates a Laplace$(a_i , 1/\epsilon)$ random variate.} To show that this mechanism does indeed satisfy DP, we suppose that $a_i=a_i^\prime$ for $i= 1, \hdots,k-1, k+1, \hdots, K$ and that $a_k^\prime = a_k-1$ (\added{i.e.\ the assumptions made in Section \ref{sec2}}). Firstly, when $b_k>a_k$: 
\begin{align}
\frac{\mathbb{P}(\mathcal{M}_L(\mathbf{a})=\mathbf{b})}{\mathbb{P}(\mathcal{M}_L(\mathbf{a}^\prime)=\mathbf{b})} &= \frac{ \exp \left(-{\epsilon |b_k - a_k|}\right)}{ \exp \left(-{\epsilon |b_k - a_k^\prime|}\right) } \nonumber \\ &= \frac{ \exp \left(-{\epsilon |b_k - a_k|}\right)}{ \exp \left(-{\epsilon |b_k - (a_k-1)|}\right) } \nonumber \\ &= \exp{ (\epsilon)} \label{laplace}. 
\end{align}
Similarly, when $a_k>b_k$, (\ref{laplace}) is equal to exp(-$\epsilon$), and when $a_k=b_k$ it is equal to exp(0). Hence the DP definition in (\ref{DP}) holds. 
\begin{example}[The Gaussian mechanism] \end{example}
A random variable $X \sim$ Normal$(\mu ,\sigma^2)$ has probability density function $f_G$:

\begin{align*}
f_G(x ; \mu ,\sigma^2)&=\frac {1}{\sigma \sqrt{2\pi}} \exp \left[ -\frac{1}{2} \left({\frac {x-\mu }{\sigma}}\right)^2 \right] 
\end{align*}
In a similar way to the Laplace mechanism, the Gaussian mechanism, say $\mathcal{M}_G$, applies Normal(0,\; $\sigma^2$) random noise to the original counts, resulting in a mechanism that satisfies $(\epsilon, \delta)$-differential privacy. Using the same assumptions and notation as previous, it follows that:   
\begin{align*}   \frac{\mathbb{P}(\mathcal{M}_G(\mathbf{a})=b)}{\mathbb{P}(\mathcal{M}_G(\mathbf{a^\prime})=b)}&= \frac{
{\frac {1}{\sigma {\sqrt {2\pi }}}} \exp \left[ {-{\frac {1}{2}}\left({\frac {b_k-a_k }{\sigma }}\right)^{2}} \right]}{
{\frac {1}{\sigma {\sqrt {2\pi }}}} \exp \left[ {-{\frac {1}{2}}\left({\frac {b_k-a_k+1 }{\sigma }}\right)^{2}} \right] } \\
&= \exp \left[ -{\frac {1}{2\sigma^2}} (2a_k-2b_k-1)\right].
\intertext{Recall that $(\epsilon, \delta)$-probabilistic DP is satisfied whenever}
\frac{1}{\exp{(\epsilon)}} &\leq \frac{\mathbb{P}(\mathcal{M}(\mathbf{a})=b)}{\mathbb{P}(\mathcal{M}(\mathbf{a^\prime})=b)}\leq \exp{(\epsilon)} \quad \text{with probability $1 - \delta$,} \intertext{which, in this instance, occurs whenever}
   -\epsilon  &\leq -{\frac {1}{2\sigma^2}} (2a_k-2b_k-1) \leq \epsilon \quad \text{with probability $1 - \delta$}.
   \end{align*}
The probability $1-\delta$ can be obtained from $\Phi$, the normal distribution's CDF \citep{Balle2018}, as $b_k \sim \text{Normal}(a_k, \sigma^2)$.
\begin{align*} 1-\delta&= \mathbb{P} (-\epsilon  \leq -{\frac {1}{2\sigma^2}} (2a_k-2b_k-1) \leq \epsilon)\\
&=\mathbb{P} (a_k -\sigma ^2 \epsilon - {1}/{2} \leq b_k \leq a_k +\sigma ^2 \epsilon - {1}/{2})\\
            &=\Phi \left( \frac{ a_k +\sigma ^2 \epsilon - {1}/{2}-a_k}{\sigma} \right) - \Phi \left( \frac{ a_k -\sigma ^2 \epsilon - {1}/{2}-a_k}{\sigma} \right)\\
            &=\Phi \left( { \sigma \epsilon - {1}/{(2\sigma)}} \right) - \Phi \left(- \sigma \epsilon - {1}/{(2\sigma)} \right)
\end{align*}
\begin{example}[Multinomial-Dirichlet synthesizer]
\end{example}
A multinomial-Dirichlet synthesis mechanism \citep{Abowd2008}, say $\mathcal{M}_{MD}$, can also yield DP guarantees. The original counts $\mathbf{a}$ can be converted to cell probabilities $\boldsymbol\pi$ simply by dividing by $n$ (the number of individuals in the data). A Dirichlet prior with concentration parameters $\boldsymbol\alpha=(\alpha_k, \alpha_2, \hdots, \alpha_K)$ is placed on $\boldsymbol\pi$ (see \cite{Abowd2008} for more on this approach). Using the same ``without loss of generality'' assumptions as previous, it follows that

\begin{align}
   \frac{\mathbb{P}(\mathcal{M}_{MD}(\mathbf{a})=\mathbf{b})}{\mathbb{P}(\mathcal{M}_{MD}(\mathbf{a}^\prime)=\mathbf{b})}&= \frac{\Gamma(b_k + a_k + \alpha_k)}{\Gamma(a_k + \alpha_k) } \cdot \frac{\Gamma(a_k^\prime + \alpha_k) }{\Gamma(b_k + a_k^\prime + \alpha_k)} \nonumber \\
  &= \frac{\Gamma(b_k + a_k + \alpha_k)}{\Gamma(a_k + \alpha_k) } \cdot \frac{\Gamma(a_k-1 + \alpha_k) }{\Gamma(b_k + a_k-1 + \alpha_k)} \nonumber \\
  &= \frac{b_k + a_k-1 + \alpha_k}{a_k-1 + \alpha_k}.  \label{eqn} 
    \intertext{Recall again that DP is satisfied whenever}
     \frac{1}{\exp{(\epsilon)}} &\leq \frac{\mathbb{P}(\mathcal{M}(\mathbf{a})=\mathbf{b})}{\mathbb{P}(\mathcal{M}(\mathbf{a}^\prime)=\mathbf{b})}  \leq {\exp{(\epsilon)}}. \nonumber
  \intertext{As the expression in (\ref{eqn}) is always greater than or equal to one, and hence always greater than 1/exp($\epsilon$), DP is satisfied whenever}
    \frac{b_k + a_k-1 + \alpha_k}{a_k-1 + \alpha_k} &\leq {\exp{(\epsilon)}}. \nonumber 
    \intertext{As $a_k \geq1$ and $b_k \leq n$, this simplifies to}
     \frac{n + \alpha_k}{ \alpha_k}  &\leq {\exp{(\epsilon)}} \quad \Rightarrow \quad
 \alpha_k \geq \frac{n}{\exp{(\epsilon)} -1}. \nonumber
   \intertext{Considering all counts $a_1, \hdots, a_K $ gives that DP is satisfied whenever}
 \text{max}_i \alpha_i &\geq \frac{n}{\exp{(\epsilon)} -1}, \quad \text{a result from \cite{machanavajjhala2008privacy}.} \nonumber
\end{align}
\section{Satisfying $(\epsilon, \delta)$-probabilistic DP with a Poisson synthesis mechanism} \label{sec4}
When using saturated count models to synthesize contingency tables, as set out in \cite{Jackson2021}, a count distribution, e.g.\ the Poisson, applies noise to original counts. We assume that a constant pseudocount $\alpha>0$ is added to every element of $\mathbf{a}$ (i.e.\ to \textit{all} original counts, not just to zero counts as in \cite{Jackson2021}), which opens up the possibility that original counts of zero can be synthesized to non-zeros. When using the Poisson we apply the following mechanism, which we denote by $\mathcal{M}_P$, to obtain a set of synthetic counts:
\begin{align}
    {b}_i \mid {a}_i, \alpha &\sim \text{Poisson}(a_i+\alpha), \quad i=1,\hdots,K, \nonumber  \\
    \text{i.e.} \quad \mathbb{P}(\mathcal{M}_P(a_i)=b_i) &= \frac{ \exp{(-a_i-\alpha)}(a_i+\alpha)^{b_i}}{b_i!}, \quad i=1,\hdots,K. \nonumber \\
     \intertext{Supposing once again that $\mathbf{a}$ and $\mathbf{a}^\prime$ differ in their $k$th element only, we have:}
  \frac{\mathbb{P}(\mathcal{M}_{P}(\mathbf{a})=\mathbf{b})}{\mathbb{P}(\mathcal{M}_{P}(\mathbf{a}^\prime)=\mathbf{b})}   &= \exp{(-1)} \bigg( \frac{a_k+\alpha}{a_k-1+\alpha} \bigg)^{b_k}.  \label{pois}
  \end{align}
  This quantity is bounded below by exp(-1), with this minimum occurring when $b_k=0$. It is unbounded above, however, as $b_k$ can take any integer up to infinity; i.e.\ the expression in (\ref{pois}) tends to infinity as $b_k$ tends to infinity. Thus $\epsilon$-DP cannot be satisfied. \par Instead, we now consider the $(\epsilon, \delta)$-probabilistic DP relaxation, first considering the left-hand inequality of the DP definition (Def. \ref{def1}):
     \begin{align*}
             \frac{1}{\exp{(\epsilon)}}  \leq& \; \frac{\mathbb{P}(\mathcal{M}_{P}(\mathbf{a})=\mathbf{b})}{\mathbb{P}(\mathcal{M}_{P}(\mathbf{a}^\prime)=\mathbf{b})} \quad \Rightarrow \quad b_k \geq \frac{1-\epsilon}{\text{log}\left( \frac{a_k+\alpha}{a_k-1+\alpha} \right)} .
             \end{align*}
             When $\epsilon \geq 1$, this inequality holds with probability 1. When $0 < \epsilon <1$, the probability that this inequality holds can be determined through the Poisson's CDF, since $b_k$ is a realization from a Poisson random variable. This probability is given as:
              \begin{align}
          1- F_{a_k+\alpha}^P\left[\frac{1-\epsilon}{\text{log}\left( \frac{ a_k+\alpha}{ a_k-1+\alpha} \right)}  \right], \label{lower}
  \end{align}
 where $F_{a_k+\alpha}^P$ is the CDF of the Poisson distribution with mean $a_k+\alpha$. 
   \par We next consider the right-hand inequality of Def. \ref{def1}:
             \begin{align*}
              \frac{\mathbb{P}(\mathcal{M}_{P}(\mathbf{a})=\mathbf{b})}{\mathbb{P}(\mathcal{M}_{P}(\mathbf{a}^\prime)=\mathbf{b})} \leq& \; {\exp{(\epsilon)}} \quad \Rightarrow \quad b_k \leq \frac{1+\epsilon}{\text{log}\left( \frac{a_k+\alpha}{a_k-1+\alpha} \right)} .
             \end{align*}
             For all $\epsilon$, this inequality holds with probability
               \begin{align}
           F_{a_k+\alpha}^P \left[\frac{1+\epsilon}{\text{log}\left( \frac{ a_k+\alpha}{ a_k-1+\alpha} \right)}  \right]. \label{upper}
  \end{align}
  Recall that in $(\epsilon, \delta)$-probabilistic DP, $1-\delta$ is the probability that DP is satisfied, i.e.\ the probability that both inequalities hold. A non-trivial question when $0 < \epsilon <1$ is how to combine the probabilities given in (\ref{lower}) and (\ref{upper}) and hence compute $\delta$? This is an area of future research. \par
\begin{figure}
    \centering
    \includegraphics[width=10cm]{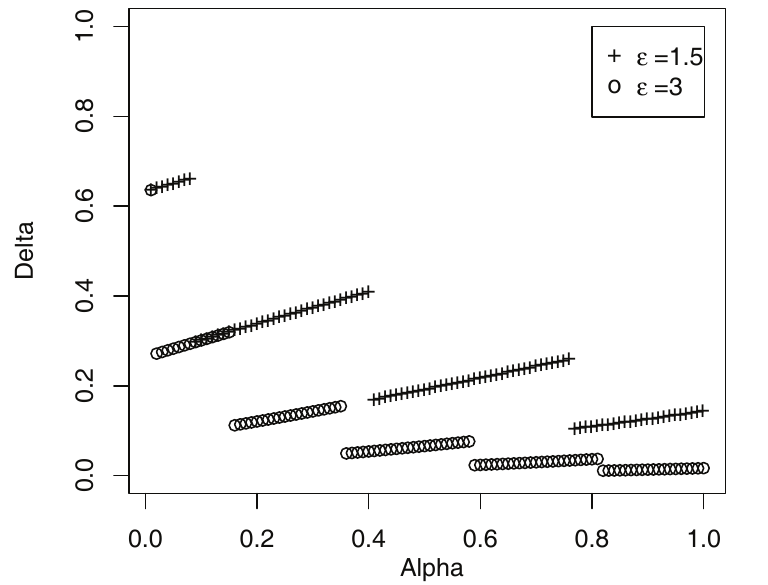}
    \caption{The relationship between $\alpha$ and $\delta$ in the Poisson synthesis mechanism for $\epsilon=1.5$ and $\epsilon=3$.}
    \label{fig1}
\end{figure}

 
  When $\epsilon>1$, however, the left-hand inequality of Def. \ref{def1} always holds, thus we need only focus on (\ref{upper}). Although non-trivial for any $\epsilon\geq1$  and $\alpha>0$, (\ref{upper}) is minimised when $a_k=1$ (when $a_k^\prime=0$). Note, a formal proof has been omitted here but extensive empirical simulation results have been undertaken. Thus,  
  \begin{align}
      1- \delta = F_{1+\alpha}^P \left[\frac{1+\epsilon}{\text{log}\left( \frac{ 1+\alpha}{ \alpha} \right)}  \right]. \label{ex1}
  \end{align}
  This also demonstrates the role of $\alpha$ as a tuning parameter for risk. In general, a larger $\alpha$ value corresponds to a lower $\delta$ value. Yet $\delta$ is not a decreasing function of $\alpha$. For a very brief explanation, this is because increasing $\alpha$ increases the value of the expression inside the squared bracket in (\ref{ex1}), but it also increases the mean of the Poisson random variable from which a synthetic count is drawn. Figure \ref{fig1} illustrates the nature of the relationship between $\alpha$ and $\delta$ for different values of $\epsilon$. For example, setting $\alpha=0.1$ satisfies approximately (3,0.3)-probabilistic DP and (1.5,0.6)-probabilistic DP. \par
In contingency tables where there are no zero counts, a $(\epsilon, \delta)$-DP guarantee can be obtained when $\alpha=0$. In this instance, $\delta$ is determined by the smallest original count, i.e.:
\begin{align}
      1-\delta= F_{a_i+\alpha}^P \left[\frac{1+\epsilon}{\text{log}\left( \frac{\text{min}_i a_i+1}{\text{min}_i a_i} \right)}  \right]. \label{ex2}
  \end{align}
 In a sense, in this example we have violated the traditional $(\epsilon, \delta)$-probabilistic DP definition given in (\ref{def3}) because $\delta$ is dependent on a particular set of original counts $\mathbf{a}$ -- not all original counts. \par
We can easily replace the Poisson with any other count distribution (e.g.\ the negative binomial, Poisson inverse-Gaussian, Delaporte, Sichel, etc.), which of course would lead to a different expression for the ratio in (\ref{pois}).  
\section{An empirical example} \label{sec5}
\subsection{The English School Census administrative database}
The English School Census (ESC) is a large administrative database belonging to the UK's Department for Education (DfE), which holds information about pupils attending state-funded schools in the UK. Owing to the presence of sensitive data, strict privacy guarantees would be required for data from the ESC to be made available to researchers. There is therefore great appeal to DP-type approaches, where more formal guarantees of privacy can be obtained. \par 
Access to the real ESC data is currently restricted, even for the sake of demonstrating the effectiveness of privacy methods. For this reason, staff at the Office for National Statistics (ONS) created a substitute data set using publicly-available data sources, such as published ESC data and 2011 UK census data. A key feature of this data set, ESC$_\text{rep}$, is that it replicates some of the statistical properties present in the actual ESC. We take a subset of this data which has approximately $8 \times 10^6$ individuals (rows) and 5 categorical variables (columns). As all variables are categorical, the data set can be expressed as a contingency table with around $3.5 \times 10^6$ cells. More information about the data set -- as well as the data set itself -- is available at \cite{Blanchard2022}.
\subsection{Applying the Poisson synthesis mechanism}
We now apply the Poisson synthesis mechanism to the ESC$_\text{rep}$ data, considering different values of $\alpha$, and considering $\epsilon>1$ values. \par
\begin{figure}
    \centering
    \includegraphics[width=11.5cm]{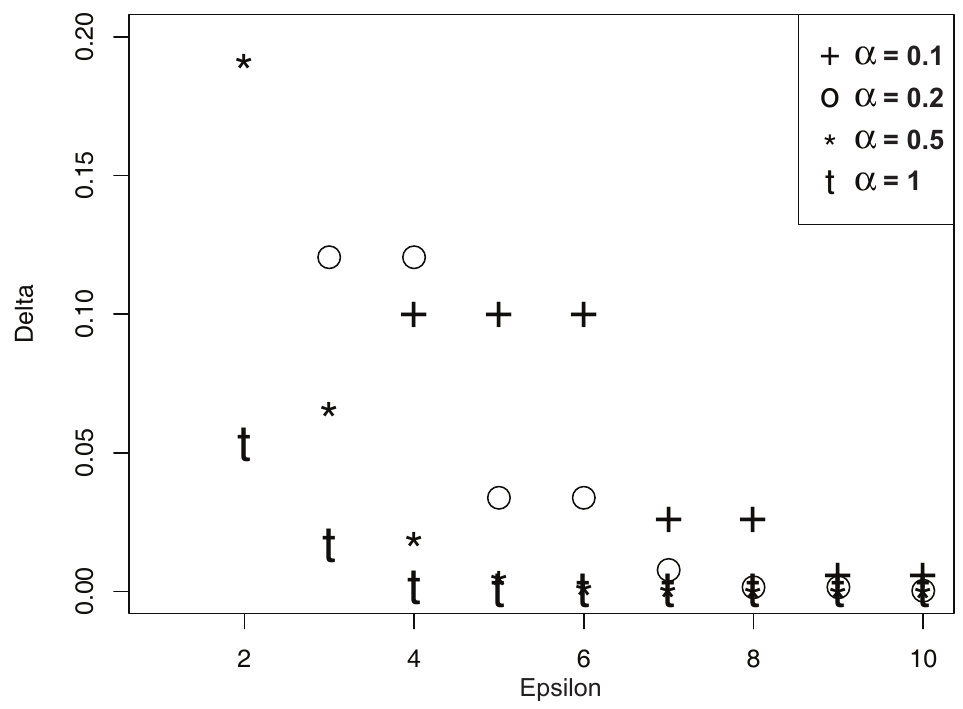}
    \caption{Combinations of $\delta$ such that $(\epsilon, \delta)$-probabilistic DP is achieved when the Poisson is used, for various max$_i a_i$ and $\epsilon$ equal to 1.5, 2, 2.5 and 3.}
    \label{fig2}
\end{figure}

 Figure \ref{fig2} gives combinations of $(\epsilon, \delta)$ values that can be achieved for the ESC$_\text{rep}$ data when using $\alpha$ values of 0.1, 0.2, 0.5 and 1. For example, when $\epsilon=2$, an $\alpha$ value of 1 is required to obtain a $\delta$ value of 0.05; when $\alpha=0.1$, a $\delta$ value of 0.05, is obtained only for $\epsilon$ values greater than 6. \par
\begin{figure}
    \centering
    \includegraphics[width=11.5cm]{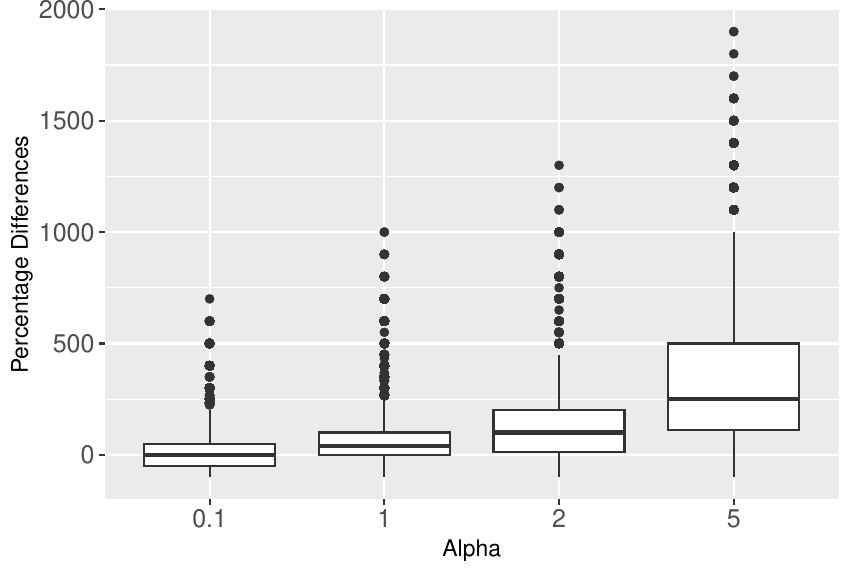}
    \caption{For different values of $\alpha$, boxplots showing percentage differences between original and synthetic counts (utility) for original counts in the range 1--10.}
    \label{fig3}
\end{figure}

 DP methods,in general are known to have a detrimental effect on utility. To gain a simple insight into general utility \citep{Snoke2018}, the boxplots in Figure \ref{fig3} compare the percentage differences between original and synthetic counts for various values of $\alpha$, and for original counts between 1 and 10. Unsurprisingly, increasing $\alpha$ increasing the percentage differences, i.e.\ has an adverse effect on utility. This loss of utility is more magnified in specific analyses, especially when the analyst wishes to quantify uncertainty. 
\section{Discussion} \label{sec6}
To summarise, in this paper we have shown how to obtain $(\epsilon, \delta)$-DP guarantees when using a Poisson synthesis mechanism to protect the privacy of counts in contingency tables. \added{For a given $\epsilon>1$, the corresponding value of $\delta$ that is achievable with the Poisson is relatively high; much higher than that which is achievable with other DP mechanisms.} Going forward, we believe other count distributions, such as the negative binomial, are likely to be more favourable (i.e.\ will give better utility results), while also providing the same DP-type risk guarantees, because such distributions would introduce further tuning parameters in addition to $\alpha$. Previous work suggests that such tuning parameter apply noise in a more efficient fashion \citep{Jackson2022}. These tuning parameters could be set to obtain certain $\epsilon$ or $\delta$ values. 

We end with an interesting note in relation to DP. Somewhat counterintuitively, the reason why multinomial-based synthesis mechanisms (e.g.\ the multinomial Dirichlet synthesizer) can satisfy $\epsilon$-DP -- but the Poisson cannot -- is because with multinomial mechanisms have a maximum synthetic count that any original count can take, namely $n$. With count distributions, any original count can be synthesized to any non-negative integer. To help explain why this causes the DP definition to fail, recall that, with contingency tables, DP definitions effectively assume that the intruder is trying to locate the cell to which
just one individual belongs; i.e.\ in the intruder's data set one, and only one, cell count is one less than it actually is. Suppose that a particular count in the intruder's data set is equal to 1, but that the corresponding synthetic count -- generated by simulating from the Poisson with $\alpha=0$ -- has a count of 5. It is 11.7 times more likely that this synthetic count originated from a cell with a count of 2 than from a count of 1, therefore the intruder can infer that that particular cell is a likely origin of the target. It is interesting therefore that, with DP, disclosure risk is deemed to be at its greatest when the scope for potential movement between original and synthetic counts is at its greatest. This largely goes against the objectives of traditional SDC methods, which typically reduce risk by increasing the divergence from the original counts.
\bibliography{main}
\end{document}